\documentclass[twocolumn,  trackchanges]{aastex701}
\usepackage{color}     
\usepackage[utf8]{inputenc}

\newcommand{\Fig}[1]{Figure~\ref{#1}}
\newcommand{\Figs}[2]{Figures~\ref{#1} and \ref{#2}}

\newcommand{\Sec}[1]{Section~\ref{#1}}

\newcommand{\Secs}[2]{Sections~\ref{#1} and \ref{#2}}

\newcommand{\degree}{^{\circ}}

\begin{document}

\title{Backtracking Bipolar Magnetic Regions to their emergence: Two groups and their implication in the tilt measurements}

\author[orcid=0000-0001-7036-2902,sname='Sreedevi']{Anu Sreedevi}
\affiliation{Department of Physics, Indian Institute of Technology (Banaras Hindu University), Varanasi, 221005, India}
\email[show]{anubsreedevi.rs.phy20@itbhu.ac.in}  

\author[orcid=0000-0002-8883-3562, sname='Karak']{Bidya Binay Karak} 
\affiliation{Department of Physics, Indian Institute of Technology (Banaras Hindu University), Varanasi, 221005, India}
\email[show]{karak.phy@iitbhu.ac.in}

\author[orcid=0000-0003-3191-4625, sname='Jha']{Bibhuti Kumar Jha}
\affiliation{Southwest Research Institute, Boulder, CO 80302, USA}
\email{bibhuti.jha@swri.org}

\author[orcid=0009-0003-7500-7258, sname='Gupta']{Rambahadur Gupta}
\affiliation{Department of Physics, Indian Institute of Technology (Banaras Hindu University), Varanasi, 221005, India}
\email{rambahadurgupta.rs.phy23@itbhu.ac.in}

\author[orcid=0000-0003-4653-6823,sname='Banerjee']{Dipankar Banerjee}
\affiliation{Indian Institute of Astrophysics, Koramangala, Bangalore 560034, India}
\affiliation{Center of Excellence in Space Sciences India, IISER Kolkata, Mohanpur 741246, West Bengal, India}
\email{dipu@iist.ac.in}

\begin{abstract}

Bipolar Magnetic Regions (BMRs) that appear on the solar photosphere are surface manifestations of the Sun’s internal magnetic field. With modern observations and continuous data streams, the study of BMRs has moved from manual sunspot catalogs to automated detection and tracking methods. In this work, we present an additional module to the existing BMR tracking algorithm, AutoTAB, that focuses on identifying emerging signatures of BMRs. Specifically, for regions newly detected on the solar disk, this module backtracks the BMRs to their point of emergence. From a total of about 12,000 BMRs identified by AutoTAB, we successfully backtracked 3,080 cases. Within this backtracked sample, we find two distinct populations. One group shows the expected behavior of emerging regions, in which the magnetic flux increases significantly during the emerging phase. The other group consists of BMRs whose flux, however, does not exhibit substantial growth during their evolution, the instances where our algorithm fails to capture the initial emergence of the BMRs. We classify these as “discarded” BMRs and examine their statistical properties separately. Our analysis shows that these discarded BMRs do not display any preferred tilt angle distribution and do not show systematic latitudinal tilt dependence, in contrast to the trends typically associated with emerging BMRs. This indicates that including such regions in statistical studies of BMR properties can distort or mask the underlying physical characteristics. We therefore emphasise the importance of excluding the discarded population from the whole dataset when analysing the statistical behavior of BMRs.

\end{abstract}


\section{Introduction} 

Isolated, intense magnetic field regions on the solar photosphere manifest as dark spots, often appearing in
pairs or groups, in white-light images. When observed in magnetically sensitive spectral lines, these sunspot pairs correspond to opposite polarities of a more general magnetic structure known as bipolar magnetic regions \citep[BMRs;][]{H08, H1991a, H1991}. Continuous records of sunspots have been maintained since the early 1600s, following the popularization of telescopes, while systematic observations of their magnetic counterparts date back to the late 1900s with the advent of magnetographs \citep{SG1999}. Owing to the long standing observational record, these features have been studied for over a century. 
Since sunspots are the primary sites of solar eruptive events, they serve as a key indicator of solar activity \citep{S2009}. Consequently, 
study of 
the evolution and properties of individual BMRs, is essential for understanding and improving the predictability of the solar events 
and solar cycle \citep{Petrovay20, K2023}.

The cataloging of solar magnetic regions has traditionally relied on human observers and sunspot records. In the last decade, data from various ground-based observatories have been digitized and made publicly available, leading to the development of several comprehensive, cross-calibrated sunspot catalogs \citep{AV2020}. 
The advent of high-cadence satellite observations in recent years (Michelson Doppler Imager (MDI) aboard Solar and Heliospheric Observatory (SOHO) and the Helioseismic and Magnetic Imager (HMI) aboard Solar Dynamic Observatory (SDO)) demand a better way of studying and cataloging BMRs and out of the reach of human observers. This helped in the development of automatic techniques to detect and keep track of the evolving BMRs and sunspots through their nearside and farside evolution. Starting initially with automatic detection, 
\citet{SK2008} and \citet{SK2012} introduced a method for automatically isolating strictly bipolar structures in line-of-sight magnetograms through a 
flux-based threshold, a technique later adapted to HMI data by \citet{JK2020} and \citet{SJ2023}. Subsequently, \citet{TP2014} developed an automated algorithm for identifying penumbrae and umbrae from daily HMI white-light images. Threshold-based approaches have since improved through morphological operations such as opening/closing and region growth techniques. However, analyses based on such datasets tend to emphasize long-lived BMRs, as they persist longer in the dataset. This bias can be mitigated by tracking BMRs throughout their evolution, allowing researchers to represent a given region at any desired stage and led to development of various tracking algorithms of BMRs. The Solar Monitor Active Region Tracking \citep[SMART]{HG2011} algorithm became one of the early tools for real-time detection and tracking, aimed at supporting solar eruptive event prediction. Optimzed for space weather prediction, they overlook small features. The Bipolar Active Region Detection \citep[BARD]{JW2016} system and dataset from \citet{WJ2024}  employs a similar detection framework but employs a dual-maximum flux-weighted overlap method for feature association and incorporates human oversight for complicated pairing to ensure efficiency. Similar datasets are provided by Space-Weather MDI Active Region Patches \citep[SMARP]{BS2014} and Space-Weather HMI Active Region Patches \citep[SHARP]{BW2021} extend the tracking capability by providing magnetogram-based maps of active regions identified in MDI and HMI observations. However, active regions in SMARPs and SHARPs are not necessarily true BMRs, as they may lack sufficient flux balance. More recently, the rapid rise of machine learning has brought AI-based detection and tracking models into widespread use, offering speed and adaptability beyond traditional methods \citep{GY2024}. Yet, a fundamental challenge remains: most existing techniques, whether manual, threshold-based, tracking-based, or AI-driven still fail to capture the very early stages of BMR emergence and rarely differentiate between the emerging BMRs  
and the non-emerging phase of the 
BMRs or the bipolar  
structures arising from fragmentation of decaying BMRs.

In this study, we aim to identify and track BMRs from their earliest stages of emergence, distinguishing 
BMRs from emerging/decaying phase and fragmented regions, and examining their properties.
We shall show that a significant fraction of BMRs detected and tracked in the LOS magnetograms do not correspond to true emerging phases of BMRs, and their properties (e.g., the tilt angle) are quite different from what one expects for true emerging BMRs. We shall show that it is essential to isolate these regions from the primary data before studying the properties of true emerging BMRs.  

The paper is organized as follows. \Secs{sec:data}{sec:methodology} respectively describe the data sets used and the working of our automatic detection algorithm.  \Sec{sec:results} details the methodology for classifying BMRs as truly emerging regions and the others and analyzes their respective properties. Finally, \Sec{sec:conclusion} summarizes our findings and present concluding remarks.

\section{Data} \label{sec:data}
For our study we use the line-of-sight (LOS) magnetogram data from two instruments, namely Michelson Doppler Imager \citep[MDI;][]{SB1995} onboard SOHO and the Helioseismic and Magnetic Imager \citep[HMI;][]{SS2012} onboard SDO. MDI provides LOS magnetograms with a spatial resolution of approximately 4 arcseconds and a cadence of 96 minutes from 1996 to 2011, whereas HMI provides the same with an exceptional spatial and temporal resolution of 1 arcsecond every 45 seconds and 12 minutes since 2010. The HMI LOS magnetograms are rebinned to the MDI resolution and scaled by a factor of 1.4 for consistency. In this study, we use the full dataset spanning 1996–2024. 

\section{Methodology} \label{sec:methodology}
For the objective mentioned in the Introduction, we employ our algorithm in the LOS magnetograms. 
The algorithm itself operates in three stages: (1) automatic detection, (2) automatic tracking, and (3) automatic backtracking to identify emergence. The first two stages are implemented as part of the Automatic Tracking Algorithm for Bipolar Magnetic Regions \citep[AutoTAB;][]{SJ2023}. Although the details of AutoTAB are provided in the referenced article, the algorithm has since undergone several improvements. Therefore, we start with a brief description of AutoTAB, followed by the backtracking procedure.

 \subsection{AutoTAB}
AutoTAB is an automated state-of-the-art tool designed to track the evolution of BMRs during its passage across the visible solar disk as seen from the Sun-Earth line. 
The algorithm constitutes two major steps: the detection and tracking of BMRs. 

We begin by correcting the LOS magnetic field measurements for projection effects. This correction is done by dividing the magnetic field value at each pixel by the cosine of its heliocentric angle, the angle between the line of sight and the normal to the solar surface at that pixel’s location. Next, we smooth the image using a boxcar kernel of the size of 11 pixels (22~arcsec) and apply an adaptive threshold based on the average magnetic field strength in the image \citep{SK2012}. This step helps us identify only those pixels where the magnetic field is stronger than the threshold.

To ensure the detected regions represent true bipolar magnetic regions (BMRs) and not just unbalanced patches of flux, we apply a flux balance condition at the end of the detection process. This condition ensures that the total positive and negative magnetic flux in a region are reasonably balanced. The flux balance ratio is defined as:
\begin{equation}
    r = \frac{|\Phi_{+}| -|\Phi_{-}|}{|\Phi_{+}| +|\Phi_{-}|}, \label{eq:1}
\end{equation}
where $\Phi_+$ and $\Phi_-$ are the total positive and negative magnetic fluxes in the region, respectively. A region is accepted as a BMR only if the flux balance ratio satisfies $r \leq 0.4$. Once the BMRs are identified, we extract the parameters that will be used to analyze this data set. These parameters are flux-weighted centroids location, latitude ($\lambda$), longitude ($\phi$) and other relevant parameters such as the maximum field strength in a BMR ($B_{\rm max}$) and total unsigned magnetic flux ($\Phi$). Finally, the identified region masks are stored for tracking in the next step.

Once detection is complete for all magnetogram data, the stored binary masks undergoes pre-processing steps of image opening operation using an adaptive circular kernels to extract the boundaries of the BMR. This modified binary mask along with the corresponding magnetograms, are used to track the evolution of BMRs. A detailed description of the tracking algorithm can be found in \citet{SJ2023}; here, we summarise the main steps.

The tracking process begins by isolating a BMR in its precomputed binary mask. Next, we estimate how long this BMR can be observed before it reaches the western limb of the solar disk. This is done by calculating its heliographic longitude ($\phi$) and latitude ($\theta$) and using the known differential rotation profile \citep{HH1990} at its latitude to estimate the time for the BMR to reach the limb. Using this time estimate and the cadence of the observations (96 minutes in our analysis), we determine the maximum number of future observations in which this BMR can potentially be tracked. For each of these future time steps, we predict the expected location of the BMR using the differential rotation profile, followed by applying a feature association technique to find a matching structure.

The feature association algorithm checks that the change in area (i.e., the number of pixels) between two consecutive observations remains within a range of 80\% to 300\%. In cases where data is missing, this condition is made stricter to ensure robust tracking of the same BMR. The tracking continues either until the estimated time of limb passage is reached or until the BMR fails to meet the association criteria described above. Once a BMR is identified and tracked successfully, all its information is stored in a catalog. This catalog includes the unique AutoTAB-ID assigned to each BMR, along with its temporal evolution throughout the tracking period \citep{SJ2023}. The current AutoTAB catalog contains information on roughly 12,000 unique BMRs, and the backtracking algorithm is built based on this catalog.

\subsection{Backtracking of BMRs identified by AutoTAB}

Upon careful analysis of the BMRs in the AutoTAB catalog, we observed that the AutoTAB missed the initial emergence phase of the BMRs. This is primarily due to AutoTAB's stringent detection criteria, which prioritize decent flux balance between the polarities, a condition usually met at a later stage in the BMR evolution. To address this limitation, we developed a backtracking algorithm that traces each detected BMR backward in time, starting from its first appearance in the AutoTAB catalog, given the limitation of the data associated with the cadence. Our backtracking algorithm is as follows.

Once a BMR is identified by AutoTAB, we extract the region binary mask (saved during the detection) and the corresponding full-disk magnetogram (projection effects corrected) at the time of its first detection (referred as $T_0$ from now on). The region mask is expanded into a rectangular region by extending 9 pixels (18 arcsec) outward from the boundaries of the detected region, matching the bounding-box criteria used during the AutoTAB detection phase. We restricted our analysis to BMRs whose first detections occur at heliographic longitudes $ \geq -35^\circ$, ensuring sufficient preceding observations to trace the region’s emergence prior to AutoTAB's initial detection. At the time of first detection ($T_0$), we calculate three physical quantities within the extracted subregion: (a) $R(T_0)$, the ratio of pixels with absolute magnetic field strength exceeding 100~G to the total number of pixels in the box; (b) $\Phi(T_0)$, the total unsigned magnetic flux of the region; and (c) $B(T_0)$, the flux density, calculated as the unsigned flux divided by the area of the box in pixels. Together, these characterize the magnetic strength and coherence of the BMR at the point of first detection.

\begin{figure*}
\centering
    \includegraphics[width=\textwidth]{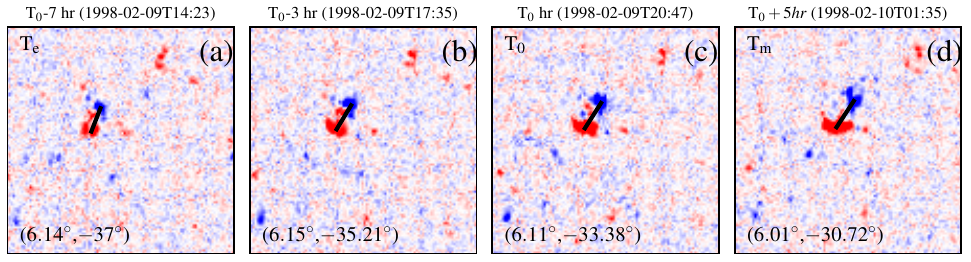}
    \includegraphics[width=\textwidth]{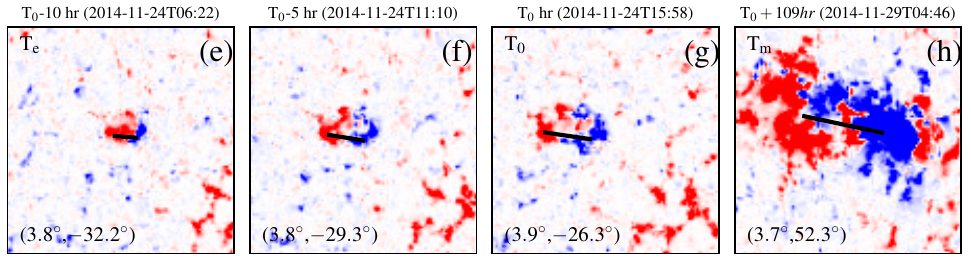}
\caption{
Evolution of two emerging/growing BMRs.
(a) to (d) The BMR at the times of initial detection ($T_{e}$), at the middle of the backtracking phase,  the starting of our backtracking phase ($T_0$), and at its maximum flux ($T_m$). Note the time sequence: $T_e \rightarrow T_0  \rightarrow T_m$.  
(e--h) 
The same time frames but for a different BMR. 
Numbers in brackets on each panel denote the mean latitude and longitude of the region. The mean latitude and longitude of the region are printed in brackets on each panel. LOS field is saturated at 1.5~kG in all panels. The line in each panel connects the flux-weighted centroids of BMR's poles. 
}
	\label{fig:bt_eg}
\end{figure*}

Before initiating the backtracking procedure, we apply a preliminary filtering step to remove potentially unreliable detections. Regions with bounding boxes larger than 100 pixels in either width or height are excluded, as such cases are typically indicative of faulty detections carried over from AutoTAB and are unlikely to correspond to genuine BMR emergence. In addition, regions with a $B(T_0)$ (flux density) below a threshold of 15 are discarded, as they often represent weak, incoherent magnetic fields (e.g., salt-and-pepper noise) or decaying remnants of earlier activity rather than newly emerging BMRs. After filtering, the maximum allowable backtracking duration is estimated using the same rotational model adopted in the AutoTAB tracking algorithm. 

\begin{figure*}
	\centering
	\includegraphics[width=1.0\textwidth]{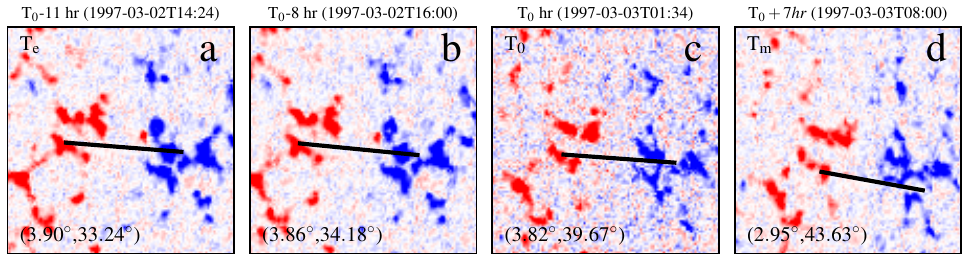}
	\includegraphics[width=1.0\textwidth]{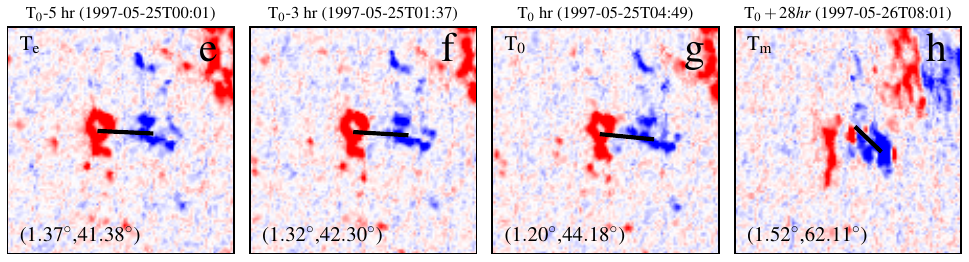}
	\includegraphics[width=1.0\textwidth]{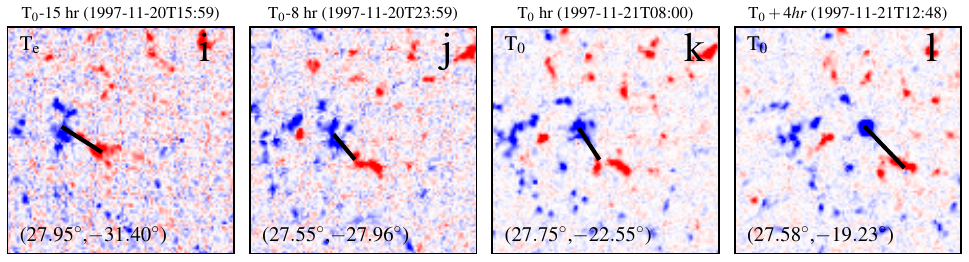}
        \includegraphics[width=1.0\textwidth]{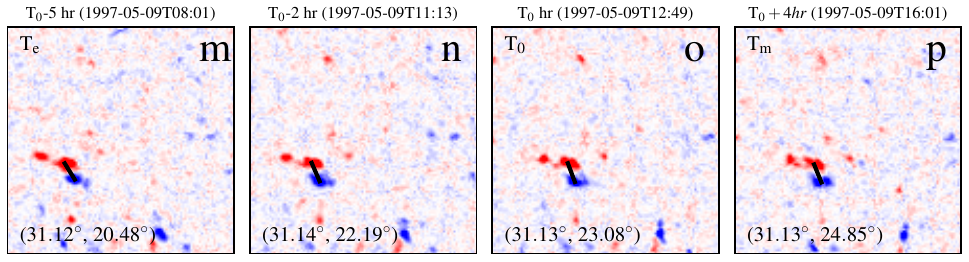}
	\caption{
    Examples of the time evolution of four discarded BMRs. 
		The figure format is the same as Figure~\ref{fig:bt_eg}, but different rows represent the time evolutions of different BMRs that fall in the same discarded BMR group (red points in Figure~\ref{fig:flux_evol}a) that were discarded in \citet{sreedevi25a}.}
	\label{sfig:bt_eg}
\end{figure*}

The algorithm then proceeds by iteratively scanning earlier magnetograms, one by one, in reverse chronological order by differentially rotating the region mask back in time to align with the expected location at previous times. If the mean longitude of the rotated region falls below $-45^\circ$, the procedure is terminated to avoid severe projection effects near the limb. At each valid step, the region’s magnetic properties are re-evaluated by calculating $R(t)$, the ratio of strong-field pixels (above 100 G) to the total number of pixels in the original box, and $\Phi(t)$, the unsigned magnetic flux. If the region contains no detectable positive or negative polarity pixels, then the step is considered invalid and the algorithm proceeds to the next earlier frame without using the current one. When both polarities are present, two diagnostic ratios are computed to evaluate the evolution of the region relative to its properties at $T_0$: (a) the pixel ratio \( r_p  = \frac{R (t)}{R (T_{0})}\) and (b) the flux ratio $r_{f} = \frac{ \Phi (t)} {\Phi (T_0)}$. These ratios are used to classify the behavior of the region. If $r_{f} > 1$ and $r_{p} > 1$, the region appears to grow unrealistically during backtracking, suggesting an unreliable step. If $r_{f} \leq 0.4$ and $r_{p} \leq 0.5$, the region is interpreted as either decaying rapidly or becoming too weak to be distinguished from the background noise level; in such cases, the step is skipped. If more than five instances are skipped in a row, the backtracking process is terminated. In all other cases, the region is deemed to have successfully backtracked. These thresholds were determined through trial and error and found to be robust against small variations. The final timestamp of backtracking is now the actual emergence time, marked as $T_e$ for the rest of the paper. Backtracking is a very challenging task, as BMRs are not necessarily well flux balanced during the emergence, and the field strength remains near the quiet Sun (salt and pepper) magnetic field value. Despite this, the emergences of some BMRs remain undetected because they either appeared near the east limb, appeared on the Sun's far side, or were already in a decaying phase upon first detection on the nearside. Of 11,987 BMRs initially present in the AutoTAB catalog, we could successfully backtrack 3,080 BMRs starting from $T_0$ to $T_e$. To show how well our code captures the emerging phases of BMRs, we show the time evolution of two 
BMRs in Figure~\ref{fig:bt_eg} (four more cases in Figure~\ref{sfig:bt_eg}). As evident in this figure, the code detects the very early emergence phase of these BMRs when they show faint signatures in the magnetogram (and almost no signature in Intensity Continuum; Figure~\ref{sfig:IC}) and effectively tracks them as long as they are on the near side of the solar disk. 

\section{Results} \label{sec:results}

After backtracking, we obtain information for 3080 BMRs, covering their evolution from early emergence until either their decay or their passage beyond the west limb, whichever occurs first. The rest of our analysis is based on this catalog. It is worth noting that for some BMRs, the backtracking algorithm identifies $T_0$ as the same as $T_e$; such cases are also included in the analysis.  Building on this, we now proceed to examine how the statistical properties of these BMRs evolve over time.

\subsection{Two groups of BMRs: Emerging and Discarded BMRs}

\begin{figure*}
    \includegraphics[width=0.51\textwidth]{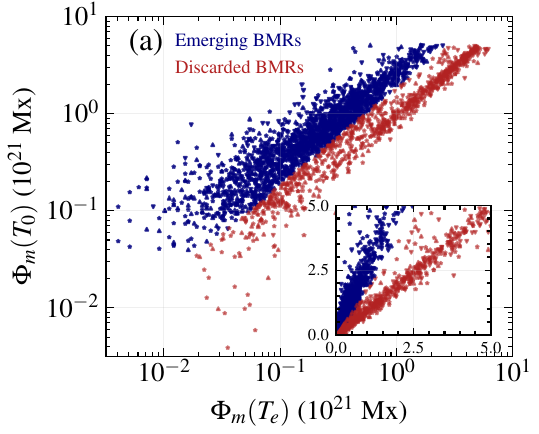}
    \includegraphics[width=0.47\textwidth]{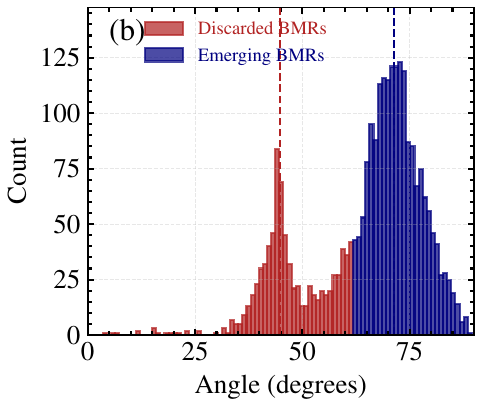}
    \includegraphics[width=0.51\textwidth]{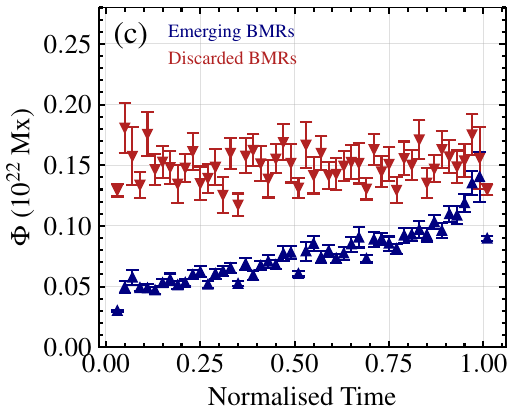}
    \includegraphics[width=0.47\textwidth]{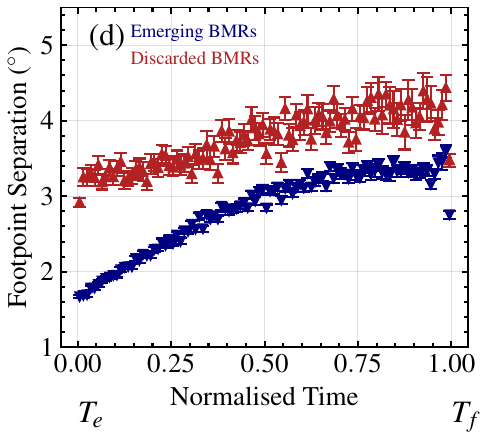}
    \caption{
    (a) Flux of BMRs ($\Phi_m$) at the start of the backtracking phase ($T_0$) versus $\Phi_m$ at the time of emergence $T_e$. The inset shows the same but in linear scales. Asterisks, upper triangles, and lower triangles, respectively, represent solar cycles 23, 24, and a part of 25. 
    Red and blue points represent the `discarded' and `emerging' BMRs, isolated based on the angle distribution.    (b) Histogram of angles measured at the origin with respect to the horizontal axis. The blue/red Gaussian curves represent the two components obtained from the Gaussian Mixture Model (GMM) fit to the angle distribution. 
    Usable BMRs are defined based on the probability of their angles falling into the upper (high-angle) cluster is $> 90\%$.  (c--d), Evolution of the flux and the footpoint separation of `discarded' and `emerging' BMRs as a function of the normalized time with a bin size of 0.01.     The time is normalized by the time period between $T_0$ and $T_e$ for (c), and the total lifetime for (d) of each BMR to bring it within 0 and 1.
    }
	\label{fig:flux_evol}
\end{figure*}

A closer look at some of the BMRs in \Figs{fig:bt_eg}{sfig:bt_eg}, we see that, in most cases, we can follow BMRs from their earliest stage through to maturity. However, some cases show little or no flux growth over time. In such cases the algorithm fails to capture the emergent phase of the BMRs, instead it captures the decaying phase of the regions or the fragmented parts of an active region.
To investigate this, we examined the change in magnetic flux between first detection ($T_e$) and a later stage ($T_0$); this helps separate genuine emergences from cases influenced by flux dispersal or reconfiguration of pre-existing fields. Figure~\ref{fig:flux_evol}a (with an inset on a linear scale) shows a scatter plot of flux at $T_e$ versus flux at $T_0$. Two clusters are evident: one along the diagonal, indicating little change in flux from the time of emergence to the later time $T_0$, and another above it, showing the expected growth of emerging BMRs \citep{svanda25}. The latter fits the conventional expectation of emerging and growing BMRs from the convection zone, while the former does not. Visual inspection confirms that most of the BMRs in the diagonal branch arise from remnant flux of old or decaying regions (Figure~\ref{sfig:bt_eg}a–d), fragmented flux from a parent BMR (Figure~\ref{sfig:bt_eg}e–h), dispersed or short-lived flux (Figure~\ref{sfig:bt_eg}i–p), or 
ephemeral regions whose (short) emerging phase is missed due to weak flux comparable to the background fields (Figure~\ref{sfig:bt_eg}m–p).  
Thus, all these are not genuine new emergences.

\begin{figure*}
       \centering
       \includegraphics[width=\textwidth]{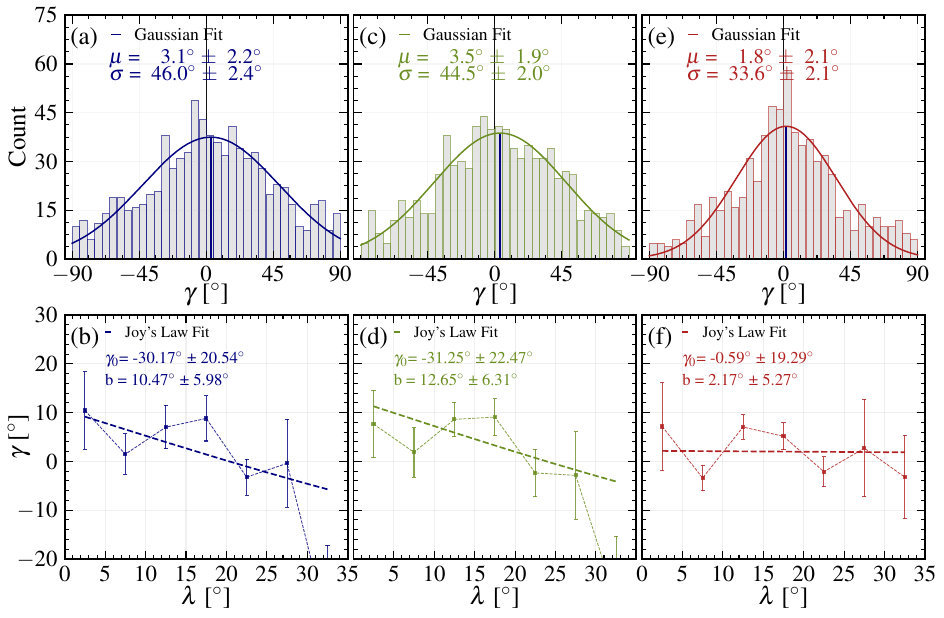}
       \caption{
       Tilts of discarded BMRs. (a) Distribution and (b) Joy's law of BMR tilt angles at $T_e$. (c) and (d) At $T_0$.  (e--f), At $T_m$. 
       Joy's law is computed by binning the tilt values in $5\degree$ latitude intervals and finding the Gaussian mean tilt for each bin. 
       The error bar is the fitted error of the Gaussian mean.} 
	\label{sfig:jlbelow}
\end{figure*}

To objectively separate the two groups, we applied a Gaussian Mixture Model (GMM) using the angle each point makes with the horizontal axis. This cleanly splits the sample into ‘emerging' and
‘discarded’ BMRs. The latter group consists of the decaying phases of BMRs or ephemeral regions whose emerging phases are missed or bipolar structures generated from the fragmentation of parent active regions (Figure~\ref{fig:flux_evol}b). Only BMRs with $\ge90\%$ probability of belonging to the growing cluster are grouped into the emerging BMR group, highlighted in blue in Figure~\ref{fig:flux_evol}a. while the red points mark the discarded ones, which make up ~27\% of the catalog. Comparison of the two groups highlights clear differences (Figure~\ref{fig:flux_evol}c–d). Growing BMRs typically start with low flux and show steady, rapid increases in unsigned flux ($\Phi_m$) from $T_e$ to $T_0$, along with consistent growth in footpoint separation. Discarded BMRs often begin with higher flux which does not increase  significantly over the period of their detection 
because many of them are the decaying BMRs and the ephemeral regions whose flux do not grow appreciably during the tracking period. 
Eventually, their flux distribution remains smaller at $T_0$ (as seen in Figure~\ref{fig:flux_evol}(a)) or in the matured phase compared to the flux distribution of emerging BMRs. The footpoint separation of these discarded BMRs also shows a distinct evolution. It begins with a large value and increases slowly and irregularly.
After manually detecting several of these discarded BMRs in the corresponding intensity Continuum data (three are shown in \Fig{sfig:IC}) we find that they are not visible as sunspots.  
All these properties strongly suggest that they arise from remnant or decaying flux rather than fresh emergence from the convection zone, consistent with our case studies. The emerging group/blue points in Figure~\ref{fig:flux_evol}, representing genuine emergences, formed the working sample analyzed separately in \citet{sreedevi25a}, where we showed robust Joy's law, starting the time of emergence $T_e$. In this article, we concentrate on the red/discarded group and how their presence affects the properties of BMRs.

\begin{figure*}
       \centering
       \includegraphics[width=\textwidth]{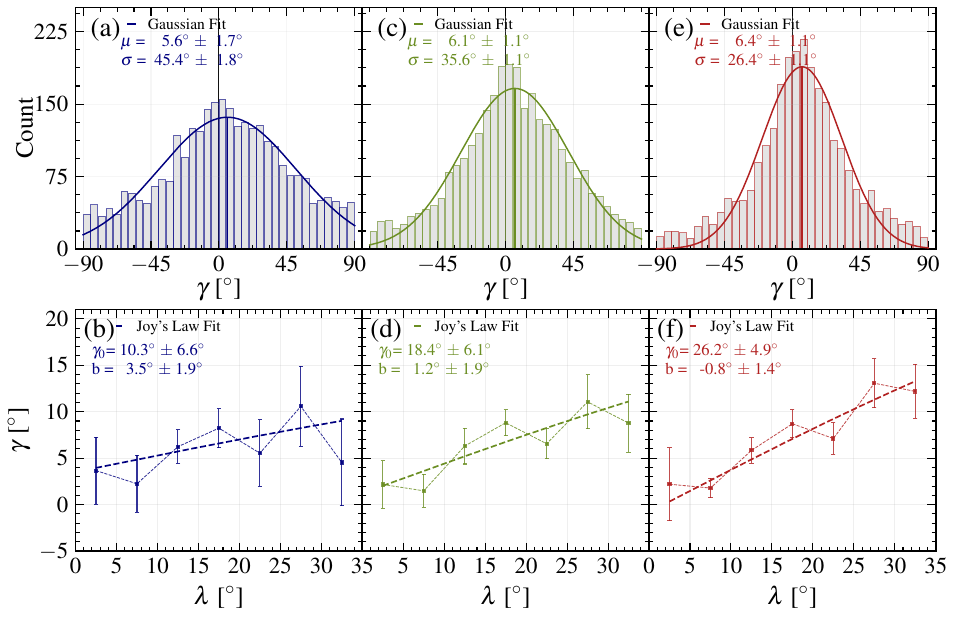}
       \caption{
       The same as Figure~\ref{sfig:jlbelow} but from all BMRs backtracked in our study i.e., without excluding the 'discarded' BMRs.
       }
	\label{sfig:jl_all}
\end{figure*}

\subsection{Tilt angle of discarded BMRs}
We compute the tilt angle $\gamma$ relative to the equator employing the usual definition of \citet{WS89}. In this, BMRs located in the northern hemisphere and exhibiting Hale–Joy orientations are assigned positive tilt values, and 
the tilt value is kept within $\pm90\degree$. To get better statistical measure, we also combine the data of two hemispheres, by reversing the sign of the tilts of southern hemisphere.

Figure \ref{sfig:jlbelow}(a, c, e) shows the tilt distribution of the BMRs that fall into the ``discarded'' groups. Across all three phases of their evolution — $T_e$, $T_0$, and $T_m$ (the matured phase of BMR)— the fitted Gaussian mean stays essentially at $0\degree$, with no clear sign of systematic tilt. 
The anti-Hale BMRs are the significant contributors in this discarded group. They contribute approximately 26\%, 25\% and 20\% at $T_e$, $T_0$, and $T_m$ respectively. These numbers in the case of the emerging BMR group are: 23\%, 13\%, and 11\%, respectively.
These anti-Hale (Joy) and Hale-Anti-Joy BMRs, both of which have negative tilts, hinder the development of the polar field, causing variations in the solar cycle and thus posing challenges in solar cycle prediction \citep{BK2023, KKS24, Dey25}.

When we look at their Joy’s law behavior in Figure \ref{sfig:jlbelow}b, d, f, we observe again a distinct picture for the discarded BMRs. At $T_e$ we see a scattered and even negative $\gamma_0$, which gradually shifts toward nearly $0\degree$ by $T_m$, but without any meaningful dependence on latitude. In other words, their tilt evolution is completely inconsistent with the expected Joy’s law trend.
This behavior of the tilt angle of discarded BMRs provides an independent check that this group of BMRs are not genuine emerging BMRs. Therefore, we need to isolate this group from the whole data set of BMRs identified in our detection code. 

\subsection{Joy's law from the combined dataset}\label{sec:JLcombined}
Since discarded BMRs do not exhibit any characteristic ‘BMR-like’ behavior, in this section we examine how combining the discarded and emerging classes affects the latitude dependence of tilt. \Fig{sfig:jl_all}(a, c, e) shows the tilt distributions of BMRs at $T_e$, $T_0$, and $T_m$ (the mature phase) respectively. At $T_e$, the fitted Gaussian mean falls at $5.6\degree \pm 1.7\degree$, but with a very wide spread. 
Over time, the spread (quantified by $\sigma$) decreases significantly while the Gaussian mean does not change much as BMRs fully emerge and reach their peak flux. 
We note that \Fig{sfig:jl_all}(e) should be the same as Figure~2(a) of \citet[]{SJ2024}. However, the smaller value of the Gaussian mean in the present data is because only a subset of BMRs could successfully be backtracked in the present study. Many high-flux BMRs that live for longer periods are not backtracked (as they either emerged near the east limb or on the far side of the Sun).

Here in Figure~\ref{sfig:jl_all}(b, d, f) we also observe a Joy’s law trend, starting from $T_e$ and 
the noise around Joy’s law decreases steadily as BMRs evolve. However, here the Joy's law trend is quite weaker than we found in our previous study \citep{sreedevi25a} excluding the discarded BMRs ($\gamma_0$ at $T_e$ was  $25.98\pm7.32$), 
demonstrating the influence of the discarded BMRs on the true signature of the BMR.
This also iterates the importance of proper identification of the tilts in the region and need of cross-validation of tilt in various datasets \citep{B2015, qin25}.
More specifically, our discarded BMRs isolated simply by observing the flux evolution will be excluded using the cross-validation of tilt from white-light measurements as done in \citet{qin25} because these discarded BMRs are mostly invisible in white-light observations.

In \cite{sreedevi25a}, we show that by separating the discarded BMRs, we find a robust Joy's law at the time of emergence on the solar surface, consistent with the thin flux tube model, which suggests that the Coriolis force causes the tilt of emerging BMRs.   If we do not isolate the discarded BMRs from the main data set as done here, then the scatter around Joy's law is considerably increased, and in a limited data set, Joy's law may not even be detected. This could be the case in previous studies \citep{SK2008}
where no evidence of Joy's law was found at the time of emergence of BMRs. 
Therefore, in any analysis of emerging BMRs, we need to remove the discarded BMRs carefully; otherwise, they may mask the actual properties of real BMRs that the Sun truly produces. 


The considerable amount of scatter in the tilt around Joy's law observed at the time of first detection ($T_e$) originates from two sources. One is due to the inability to correctly detect the very early signature of the BMR and/or incorrect identification of the flux-weighted centroids of two poles, leading to uncertainties in tilt value. This is evident in \Fig{fig:bt_wrong}(a--d), where a BMR is wrongly identified\footnote{By manually observing several BMRs we find that this type of wrong detection during back tracking phase is quite infrequent.} during the backtracking phase of our automated code and produces incorrect tilts at $t< T_e$. Another source of scatter comes from a physical origin, and it is the turbulent convection. As explained in \citet{LC2002}, during the rise of the BMR-forming flux tubes in the convection zone, they are buffeted by convective turbulence in the near-surface layer, and thus their tilts are easily disturbed. However, in the later phase, the Coriolis perturbation overpowers the short-time-scale perturbations of turbulence, which relax more rapidly. Thus, the tilt scatter reduces over the growth of the BMR as observed in our data. 
The decrease of the scatter in the tilt angle during the growth phase of BMR was also reported in the previous study of \citet{SB2020} using 153 emerging active regions. 

\section{Conclusion} \label{sec:conclusion}
Various groups have developed  algorithms for tracking the BMRs from magnetograms on the near-side, which are used to study their origins and evolutions. 
AutoTAB is one such state-of-the-art tool.
However, existing tracking approaches—whether automated or reliant on human intervention—fall short in identifying the very onset of flux emergence. To address this, we developed a backtracking algorithm that takes the first detection time ($T_0$) from AutoTAB and traces the BMRs backward to the true moment of polarity appearance, thereby defining a new emergence time ($T_e$). Applying this method to the 12,000 unique BMRs in the AutoTAB catalog, we were able to backtrack and reliably determine $T_e$ for 3080 cases. This enabled us to systematically investigate their tilt angle distributions and latitude dependence at the moment of true emergence. 

By evaluating the flux and footpoint separation evolutions between $T_e$ and $T_0$, complemented by manual observations of several BMRs, we identified two distinct BMR groups: emerging and discarded. The emerging BMRs exhibit significant flux growth, whereas the discarded ones do not show such considerable growth; instead, they consist of the decaying BMRs with nearby active regions, ephemeral regions, or fragmented bipolar flux originating from a parent active region.
We show that the discarded BMRs have almost zero mean tilt and do not show any latitudinal preference like Joy’s law, even in their mature phase. The emerging BMRs show nonzero mean tilt and a robust Joy's law even at the time of emergence on the solar surface, suggesting that they are caused by the Coriolis force, as explained by the thin flux tube model \citep{sreedevi25a}.   
If we do not isolate the discarded BMRs from the whole data set obtained from the automated code, then Joy's law trend at the time of emergence weakens as a large scatter accompanies it. 

Taken together, these results highlight a crucial implication: including discarded BMRs in statistical studies of solar active regions can severely mask the intrinsic properties of genuine, flux-emerging BMRs. To uncover the actual physical signatures of BMR emergence and their role in shaping Joy’s law, it is essential to separate these two populations.


\begin{acknowledgments}
The authors would like to thank the anonymous referee for
the comments that helped in improving the manuscript.
B.B.K. acknowledges the financial support from the Indian Space Research Organisation (project no. ISRO/RES/RAC-S/IITBHU/2024-25) and the Anusandhan National Research Foundation (ANRF) through the MATRIC program (file no. MTR/2023/000670). 
\end{acknowledgments}

\begin{contribution}
B.B.K. gave the initial idea. A.S., under the guidance of B.B.K., B.K.J., and D.B. analyzed magnetogram data, developed algorithms for BMR back tracking, and prepared the figures. R.G. analyzed Intensity Continuum data.    A.S. and B.B.K. led the manuscript writing while B.K.J. commented on the text. All authors contributed to the discussion and presentation of the results. 
\end{contribution}

\section*{Appendix}
\begin{figure*}
	\centering
        \includegraphics[width=1.0\textwidth]{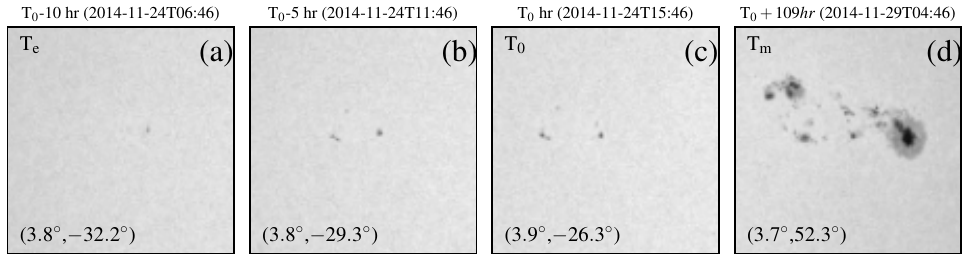}
        \includegraphics[width=1.0\textwidth]{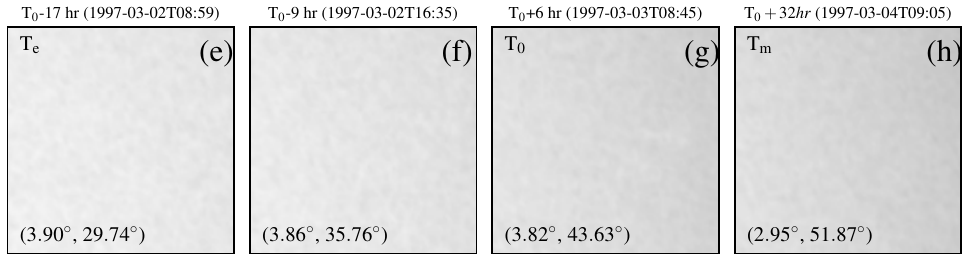}
        \includegraphics[width=1.0\textwidth]{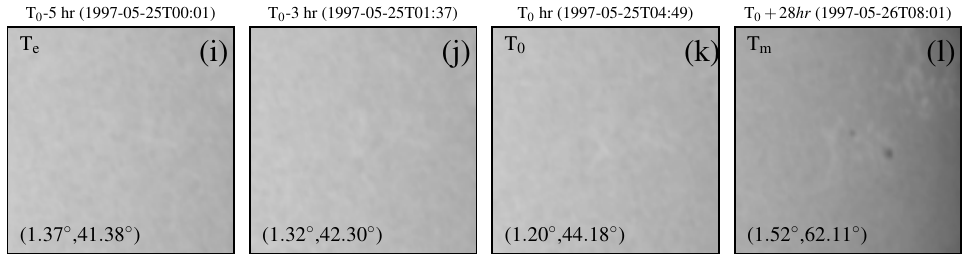}
	\caption{The time evolution of one `emerging BMR' and two `discarded BMRs' whose magnetograms are shown in Figure~\ref{fig:bt_eg}(e--h) and Figure~\ref{sfig:bt_eg}(a--d), (e--h) repectively but are now searched in corresponding (near simultaneous) intensity continuum. 
    }
	\label{sfig:IC}
\end{figure*}

\begin{figure*}
\centering
    \includegraphics[width=\textwidth]{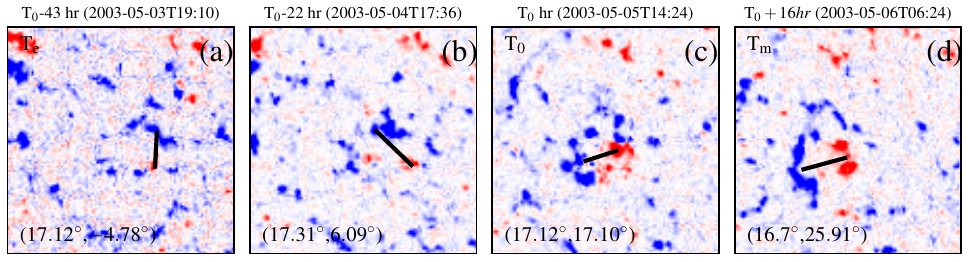}
\caption{
The same as \Fig{fig:bt_eg} but for a different BMR for which the backtracking phase ($t < T_0$) is wrongly identified. 
}
	\label{fig:bt_wrong}
\end{figure*}

This section presents additional information on the time evolution BMRs in the Intensity Continuum (\Fig{sfig:IC}) and an additional BMR in the magnetogram (\Fig{fig:bt_wrong}). 

\bibliography{sample701}{}

@ARTICLE{AV2020,
       author = {{Arlt}, Rainer and {Vaquero}, Jos{\'e} M.},
        title = "{Historical sunspot records}",
      journal = {Living Reviews in Solar Physics},
         year = 2020,
        month = dec,
       volume = {17},
       number = {1},
          eid = {1},
        pages = {1},
          doi = {10.1007/s41116-020-0023-y},
       adsurl = {https://ui.adsabs.harvard.edu/abs/2020LRSP...17....1A}
}

@ARTICLE{BK2023,
       author = {{Biswas}, Akash and {Karak}, Bidya Binay and {Kumar}, Pawan},
        title = "{Exploring the reliability of polar field rise rate as a precursor for an early prediction of solar cycle}",
      journal = {Monthly Notices of the Royal Astronomical Society},
         year = 2023,
        month = dec,
       volume = {526},
       number = {3},
        pages = {3994-4003},
          doi = {10.1093/mnras/stad2966},
archivePrefix = {arXiv},
       eprint = {2308.01155},
 primaryClass = {astro-ph.SR},
       adsurl = {https://ui.adsabs.harvard.edu/abs/2023MNRAS.526.3994B}
}

@ARTICLE{BS2014,
       author = {{Bobra}, M.~G. and {Sun}, X. and {Hoeksema}, J.~T. and {Turmon}, M. and {Liu}, Y. and {Hayashi}, K. and {Barnes}, G. and {Leka}, K.~D.},
        title = "{The Helioseismic and Magnetic Imager (HMI) Vector Magnetic Field Pipeline: SHARPs - Space-Weather HMI Active Region Patches}",
      journal = {Solar Physics},
         year = 2014,
        month = sep,
       volume = {289},
       number = {9},
        pages = {3549-3578},
          doi = {10.1007/s11207-014-0529-3},
archivePrefix = {arXiv},
       eprint = {1404.1879},
 primaryClass = {astro-ph.SR},
       adsurl = {https://ui.adsabs.harvard.edu/abs/2014SoPh..289.3549B}

}

@article{BW2021,
doi = {10.3847/1538-4365/ac1f1d},
url = {https://dx.doi.org/10.3847/1538-4365/ac1f1d},
year = {2021},
month = {sep},
publisher = {The American Astronomical Society},
volume = {256},
number = {2},
pages = {26},
author = {Monica G. Bobra and Paul J. Wright and Xudong Sun and Michael J. Turmon},
title = {SMARPs and SHARPs: Two Solar Cycles of Active Region Data},
journal = {The Astrophysical Journal Supplement Series}
}

@ARTICLE{Dey25,
       author = {{Dey}, Bidisha and {Sreedevi}, Anu and {Karak}, Bidya Binay},
        title = "{Role of Sunspot Latitude versus Tilt in Determining the Polar Field and Amplitude of the Next Cycle: Cause of the Weak Solar Cycle 20}",
      journal = {\apj},
         year = 2025,
        month = nov,
       volume = {993},
       number = {2},
          eid = {196},
        pages = {196},
          doi = {10.3847/1538-4357/ae0ccb},
archivePrefix = {arXiv},
       eprint = {2509.17146},
 primaryClass = {astro-ph.SR},
       adsurl = {https://ui.adsabs.harvard.edu/abs/2025ApJ...993..196D}
}

@ARTICLE{GY2024,
       author = {{Gong}, Long and {Yang}, Yunfei and {Feng}, Song and {Dai}, Wei and {Liang}, Bo and {Xiong}, Jianping},
        title = "{Solar Active Regions Detection and Tracking Based on Deep Learning}",
      journal = {\solphys},
         year = 2024,
        month = aug,
       volume = {299},
       number = {8},
          eid = {121},
        pages = {121},
          doi = {10.1007/s11207-024-02362-3},
       adsurl = {https://ui.adsabs.harvard.edu/abs/2024SoPh..299..121G}
}

@ARTICLE{H1991,
       author = {{Howard}, Robert F.},
        title = "{Axial Tilt Angles of Sunspot Groups}",
      journal = {Solar Physics},
         year = 1991,
        month = dec,
       volume = {136},
       number = {2},
        pages = {251-262},
          doi = {10.1007/BF00146534},
       adsurl = {https://ui.adsabs.harvard.edu/abs/1991SoPh..136..251H}
}

@ARTICLE{H08,
       author = {{Hale}, George E.},
        title = "{On the Probable Existence of a Magnetic Field in Sun-Spots}",
      journal = {The Astrophysical Journal},
         year = 1908,
        month = nov,
       volume = {28},
        pages = {315},
          doi = {10.1086/141602},
       adsurl = {https://ui.adsabs.harvard.edu/abs/1908ApJ....28..315H}
}

@ARTICLE{HG2011,
       author = {{Higgins}, P.~A. and {Gallagher}, P.~T. and {McAteer}, R.~T.~J. and {Bloomfield}, D.~S.},
        title = "{Solar magnetic feature detection and tracking for space weather monitoring}",
      journal = {Advances in Space Research},
         year = 2011,
        month = jun,
       volume = {47},
       number = {12},
        pages = {2105-2117},
          doi = {10.1016/j.asr.2010.06.024},
archivePrefix = {arXiv},
       eprint = {1006.5898},
 primaryClass = {astro-ph.SR},
       adsurl = {https://ui.adsabs.harvard.edu/abs/2011AdSpR..47.2105H}
}

@ARTICLE{HH1990,
       author = {{Howard}, R. F. and {Harvey}, J. W. and {Forgach}, S.}, 
        title = "{Solar surface velocity fields determined from small magnetic features}",
      journal = {Solar Physics},
        year = 1990,
        month = dec,
       volume = {130},
       number = {1},
        pages = {295-311},
          doi = {doi:10.1007/bf00156795},
}

@ARTICLE{JK2020,
       author = {{Jha}, Bibhuti Kumar and {Karak}, Bidya Binay and {Mandal}, Sudip and
         {Banerjee}, Dipankar},
        title = "{Magnetic Field Dependence of Bipolar Magnetic Region Tilts on the Sun: Indication of Tilt Quenching}",
      journal = {The Astrophysical Journall},
         year = 2020,
        month = jan,
       volume = {889},
       number = {1},
          eid = {L19},
        pages = {L19},
          doi = {10.3847/2041-8213/ab665c},
archivePrefix = {arXiv},
       eprint = {1912.13223},
 primaryClass = {astro-ph.SR},
       adsurl = {https://ui.adsabs.harvard.edu/abs/2020ApJ...889L..19J}
}

@INPROCEEDINGS{JW2016,
       author = {{Mu{\~n}oz-Jaramillo}, Andres and {Werginz}, Zachary and {Vargas-Acosta}, Juan Pablo and {DeLuca}, Michael and {Windmueller}, J.~C. and {Zhang}, Jie and {Longcope}, Dana and {Lamb}, Derek and {DeForest}, Craig and {Vargas-Dom{\'\i}nguez}, Santiago and {Harvey}, Jack and {Martens}, Piet},
        title = "{The best of both worlds: Using automatic detection and limited human supervision to create a homogenous magnetic catalog spanning four solar cycles}",
    booktitle = {2016 IEEE International Conference on Big Data (Big Data},
         year = 2016,
        month = dec,
        pages = {3194-3203},
          doi = {10.1109/BigData.2016.7840975},
archivePrefix = {arXiv},
       eprint = {2203.11908},
 primaryClass = {astro-ph.SR},
       adsurl = {https://ui.adsabs.harvard.edu/abs/2016bida.conf.3194M}
}

@ARTICLE{KKS24,
       author = {{Kumar}, Pawan and {Karak}, Bidya Binay and {Sreedevi}, Anu},
        title = "{Variabilities in the polar field and solar cycle due to irregular properties of bipolar magnetic regions}",
      journal = {\mnras},
         year = 2024,
        month = may,
       volume = {530},
       number = {3},
        pages = {2895-2905},
          doi = {10.1093/mnras/stae1052},
archivePrefix = {arXiv},
       eprint = {2404.10526},
 primaryClass = {astro-ph.SR},
       adsurl = {https://ui.adsabs.harvard.edu/abs/2024MNRAS.530.2895K}
}

@ARTICLE{K2023,
       author = {{Karak}, Bidya Binay},
        title = "{Models for the long-term variations of solar activity}",
      journal = {Living Reviews in Solar Physics},
         year = 2023,
        month = dec,
       volume = {20},
       number = {1},
          eid = {3},
        pages = {3},
          doi = {10.1007/s41116-023-00037-y},
archivePrefix = {arXiv},
       eprint = {2305.17188},
 primaryClass = {astro-ph.SR},
       adsurl = {https://ui.adsabs.harvard.edu/abs/2023LRSP...20....3K}
}

@ARTICLE{LC2002,
       author = {{Longcope}, Dana and {Choudhuri}, Arnab Rai},
        title = "{The Orientational Relaxation of Bipolar Active Regions}",
      journal = {Solar Physics},
         year = 2002,
        month = jan,
       volume = {205},
       number = {1},
        pages = {63-92},
          doi = {10.1023/A:1013896013842},
       adsurl = {https://ui.adsabs.harvard.edu/abs/2002SoPh..205...63L}
}

@ARTICLE{Petrovay20,
       author = {{Petrovay}, Krist{\'o}f},
        title = "{Solar cycle prediction}",
      journal = {Living Reviews in Solar Physics},
         year = 2020,
        month = mar,
       volume = {17},
       number = {1},
          eid = {2},
        pages = {2},
          doi = {10.1007/s41116-020-0022-z},
archivePrefix = {arXiv},
       eprint = {1907.02107},
 primaryClass = {astro-ph.SR},
       adsurl = {https://ui.adsabs.harvard.edu/abs/2020LRSP...17....2P}
}

@ARTICLE{qin25,
       author = {{Qin}, Lang and {Jiang}, Jie and {Wang}, Ruihui},
        title = "{Mutual Validation of Data Sets for Analyzing Tilt Angles in Solar Active Regions}",
      journal = {\apj},
         year = 2025,
        month = jun,
       volume = {986},
       number = {2},
          eid = {114},
        pages = {114},
          doi = {10.3847/1538-4357/add93a},
archivePrefix = {arXiv},
       eprint = {2502.11698},
 primaryClass = {astro-ph.SR},
       adsurl = {https://ui.adsabs.harvard.edu/abs/2025ApJ...986..114Q}
}

@ARTICLE{H1991a,
       author = {{Howard}, Robert F.},
        title = "{The Magnetic Fields of Active Regions - Part Five}",
      journal = {\solphys},
         year = 1991,
        month = mar,
       volume = {132},
       number = {1},
        pages = {49-61},
          doi = {10.1007/BF00159129},
       adsurl = {https://ui.adsabs.harvard.edu/abs/1991SoPh..132...49H}
}

@ARTICLE{B2015,
       author = {{Baranyi}, T.},
        title = "{Comparison of Debrecen and Mount Wilson/Kodaikanal sunspot group tilt angles and the Joy's law}",
      journal = {\mnras},
         year = 2015,
        month = feb,
       volume = {447},
       number = {2},
        pages = {1857-1865},
          doi = {10.1093/mnras/stu2572},
archivePrefix = {arXiv},
       eprint = {1412.1355},
 primaryClass = {astro-ph.SR},
       adsurl = {https://ui.adsabs.harvard.edu/abs/2015MNRAS.447.1857B}
}

@ARTICLE{WJ2024,
       author = {{Wang}, Ruihui and {Jiang}, Jie and {Luo}, Yukun},
        title = "{Toward a Live Homogeneous Database of Solar Active Regions Based on SOHO/MDI and SDO/HMI Synoptic Magnetograms. II. Parameters for Solar Cycle Variability}",
      journal = {\apj},
         year = 2024,
        month = aug,
       volume = {971},
       number = {1},
          eid = {110},
        pages = {110},
          doi = {10.3847/1538-4357/ad5b5f},
archivePrefix = {arXiv},
       eprint = {2405.06224},
 primaryClass = {astro-ph.SR},
       adsurl = {https://ui.adsabs.harvard.edu/abs/2024ApJ...971..110W},
}

@ARTICLE{S2009,
       author = {{Schrijver}, Carolus J.},
        title = "{Driving major solar flares and eruptions: A review}",
      journal = {Advances in Space Research},
         year = 2009,
        month = mar,
       volume = {43},
       number = {5},
        pages = {739-755},
          doi = {10.1016/j.asr.2008.11.004},
archivePrefix = {arXiv},
       eprint = {0811.0787},
 primaryClass = {astro-ph},
       adsurl = {https://ui.adsabs.harvard.edu/abs/2009AdSpR..43..739S}
}

@ARTICLE{SB1995,
   author = {{Scherrer}, P.~H. and {Bogart}, R.~S. and {Bush}, R.~I. and
        {Hoeksema}, J.~T. and {Kosovichev}, A.~G. and {Schou}, J. and
        {Rosenberg}, W. and {Springer}, L. and {Tarbell}, T.~D. and
        {Title}, A. and {Wolfson}, C.~J. and {Zayer}, I. and {MDI Engineering Team}
        },
    title = "{The Solar Oscillations Investigation - Michelson Doppler Imager}",
  journal = {Solar Physics},
     year = 1995,
    month = dec,
   volume = 162,
    pages = {129-188},
      doi = {10.1007/BF00733429},
   adsurl = {http://adsabs.harvard.edu/abs/1995SoPh..162..129S}
}

@ARTICLE{sreedevi25a,
       author = {{Sreedevi}, Anu and {Karak}, Bidya Binay and {Jha}, Bibhuti Kumar and {Gupta}, Rambahadur and {Banerjee}, Dipankar},
        title = "{Observed Joys law of Bipolar Magnetic Region tilts at the emergence supports the thin flux tube model}",
      journal = {arXiv e-prints},
         year = 2025,
        month = nov,
          eid = {arXiv:2511.03558},
        pages = {arXiv:2511.03558},
archivePrefix = {arXiv},
       eprint = {2511.03558},
 primaryClass = {astro-ph.SR},
       adsurl = {https://ui.adsabs.harvard.edu/abs/2025arXiv251103558S}
}

@ARTICLE{SB2020,
       author = {{Schunker}, H. and {Baumgartner}, C. and {Birch}, A.~C. and {Cameron}, R.~H. and {Braun}, D.~C. and {Gizon}, L.},
        title = "{Average motion of emerging solar active region polarities. II. Joy's law}",
      journal = {Astronomy \& Astrophysics},
         year = 2020,
        month = aug,
       volume = {640},
          eid = {A116},
        pages = {A116},
          doi = {10.1051/0004-6361/201937322},
archivePrefix = {arXiv},
       eprint = {2006.05565},
 primaryClass = {astro-ph.SR},
       adsurl = {https://ui.adsabs.harvard.edu/abs/2020A&A...640A.116S}
}

@ARTICLE{SJ2023,
       author = {{Sreedevi}, Anu and {Jha}, Bibhuti Kumar and {Karak}, Bidya Binay and {Banerjee}, Dipankar},
        title = "{AutoTAB: Automatic Tracking Algorithm for Bipolar Magnetic Regions}",
      journal = {The Astrophysical Journals},
         year = 2023,
        month = oct,
       volume = {268},
       number = {2},
          eid = {58},
        pages = {58},
          doi = {10.3847/1538-4365/acec47},
       adsurl = {https://ui.adsabs.harvard.edu/abs/2023ApJS..268...58S}
}

@ARTICLE{SJ2024,
       author = {{Sreedevi}, Anu and {Jha}, Bibhuti Kumar and {Karak}, Bidya Binay and {Banerjee}, Dipankar},
        title = "{Analysis of BMR Tilt from AutoTAB Catalog: Hinting toward the Thin Flux Tube Model?}",
      journal = {The Astrophysical Journal},
         year = 2024,
        month = may,
       volume = {966},
       number = {1},
          eid = {112},
        pages = {112},
          doi = {10.3847/1538-4357/ad34b8},
archivePrefix = {arXiv},
       eprint = {2403.09229},
 primaryClass = {astro-ph.SR},
       adsurl = {https://ui.adsabs.harvard.edu/abs/2024ApJ...966..112S}
}

@ARTICLE{SK2008,
       author = {{Kosovichev}, A.~G. and {Stenflo}, J.~O.},
        title = "{Tilt of Emerging Bipolar Magnetic Regions on the Sun}",
      journal = {The Astrophysical Journall},
         year = 2008,
        month = dec,
       volume = {688},
       number = {2},
        pages = {L115},
          doi = {10.1086/595619},
       adsurl = {https://ui.adsabs.harvard.edu/abs/2008ApJ...688L.115K}
}

@ARTICLE{SS2012,
       author = {{Scherrer}, P.~H. and {Schou}, J. and {Bush}, R.~I. and {Kosovichev}, A.~G. and {Bogart}, R.~S. and {Hoeksema}, J.~T. and {Liu}, Y. and {Duvall}, T.~L. and {Zhao}, J. and {Title}, A.~M. and {Schrijver}, C.~J. and {Tarbell}, T.~D. and {Tomczyk}, S.},
        title = "{The Helioseismic and Magnetic Imager (HMI) Investigation for the Solar Dynamics Observatory (SDO)}",
      journal = {Solar Physics},
         year = 2012,
        month = jan,
       volume = {275},
       number = {1-2},
        pages = {207-227},
          doi = {10.1007/s11207-011-9834-2},
       adsurl = {https://ui.adsabs.harvard.edu/abs/2012SoPh..275..207S}
}

@ARTICLE{SG1999,
       author = {{Sivaraman}, K.~R. and {Gupta}, S.~S. and {Howard}, Robert F.},
        title = "{Measurement of Kodaikanal white-light images - IV. Axial Tilt Angles of Sunspot Groups}",
      journal = {Solar Physics},
         year = 1999,
        month = oct,
       volume = {189},
       number = {1},
        pages = {69-83},
          doi = {10.1023/A:1005277515551},
       adsurl = {https://ui.adsabs.harvard.edu/abs/1999SoPh..189...69S}
}

@ARTICLE{SK2012,
       author = {{Stenflo}, J.~O. and {Kosovichev}, A.~G.},
        title = "{Bipolar Magnetic Regions on the Sun: Global Analysis of the SOHO/MDI Data Set}",
      journal = {The Astrophysical Journal},
         year = 2012,
        month = feb,
       volume = {745},
       number = {2},
          eid = {129},
        pages = {129},
          doi = {10.1088/0004-637X/745/2/129},
archivePrefix = {arXiv},
       eprint = {1112.5226},
 primaryClass = {astro-ph.SR},
       adsurl = {https://ui.adsabs.harvard.edu/abs/2012ApJ...745..129S}
}

@ARTICLE{svanda25,
       author = {{{\v{S}}vanda}, Michal and {Jur{\v{c}}{\'a}k}, Jan and {Schmassmann}, Markus},
        title = "{Average solar active region: I. Intensities, velocities, and the photospheric magnetic field}",
      journal = {\aap},
         year = 2025,
        month = aug,
       volume = {700},
          eid = {A40},
        pages = {A40},
          doi = {10.1051/0004-6361/202554440},
archivePrefix = {arXiv},
       eprint = {2506.11713},
 primaryClass = {astro-ph.SR},
       adsurl = {https://ui.adsabs.harvard.edu/abs/2025A&A...700A..40S}
}

@ARTICLE{TP2014,
   author = {{Tlatov}, A.~G. and {Pevtsov}, A.~A.},
    title = "{Bimodal Distribution of Magnetic Fields and Areas of Sunspots}",
  journal = {Solar Physics},
archivePrefix = "arXiv",
   eprint = {1308.0535},
 primaryClass = "astro-ph.SR",
     year = 2014,
    month = apr,
   volume = 289,
    pages = {1143-1152},
      doi = {10.1007/s11207-013-0382-9},
   adsurl = {https://ui.adsabs.harvard.edu/abs/2014SoPh..289.1143T}
}

@ARTICLE{WS89,
       author = {{Wang}, Y. -M. and {Sheeley}, Jr., N.~R.},
        title = "{Average Properties of Bipolar Magnetic Regions during Sunspot Cycle-21}",
      journal = {Solar Physics},
         year = 1989,
        month = mar,
       volume = {124},
       number = {1},
        pages = {81-100},
          doi = {10.1007/BF00146521},
       adsurl = {https://ui.adsabs.harvard.edu/abs/1989SoPh..124...81W}
}
\bibliographystyle{aasjournalv7}

\end{document}